\documentclass[default,iicol]{sn-jnl}

\usepackage{graphicx}%
\usepackage{multirow}%
\usepackage{amsmath,amssymb,amsfonts}%
\usepackage{amsthm}%
\usepackage{mathrsfs}%
\usepackage[title]{appendix}%
\usepackage{xcolor}%
\usepackage{textcomp}%
\usepackage{manyfoot}%
\usepackage{booktabs}%
\usepackage{algorithm}%
\usepackage{algorithmicx}%
\usepackage{algpseudocode}%
\usepackage{listings}%
\usepackage{subcaption}
\usepackage{soul}
\usepackage{cancel}

\theoremstyle{thmstyleone}%
\theoremstyle{thmstyletwo}%

\theoremstyle{thmstylethree}%

\begin{document}

\title[A Probabilistic Distance-Based Stability Quantifier for Complex Dynamical Systems]{A Probabilistic Distance-Based Stability Quantifier for Complex Dynamical Systems}


\author[1]{\fnm{Calvin} \sur{Alvares}}\email{alvarescalvin16@gmail.com}

\author*[1]{\fnm{Soumitro} \sur{Banerjee}}\email{soumitro@iiserkol.ac.in}


\affil[1]{\orgdiv{Department of Physical Sciences}, \orgname{Indian Institute of Science Education and Research}, \orgaddress{ \city{Kolkata, Mohanpur}, \postcode{741 246}, \country{India}}}




\abstract{An attractor of a dynamical system may represent the system's `desirable' state. Perturbations to the system may push the system out of the basin of attraction of the desirable attractor and into undesirable states. Hence, it is important to quantify the stability of such systems against reasonably large perturbations. {In this paper,} we introduce a distance-based measure of stability, called `basin stability bound', to characterise the stability of dynamical systems against finite perturbations. This stability measure depends on {both} the size and {the} shape of the basin of attraction of the desirable attractor. A probabilistic sampling-based approach is used to estimate basin stability bound and quantify the associated estimation error. This approach allows for the easy estimation of basin stability bound regardless of the structure of the basin of attraction and is readily applicable to high-dimensional systems. We demonstrate the merit of {the proposed} stability measure using an ecological model of the Amazon rainforest, a ship capsize model, and a power grid model.}




\maketitle

\section{Introduction}\label{sec1}
Dynamical system models of real-world phenomena often have more than one attractor \cite{feudel2018multistability}, one of which can be an `attractor at infinity'. Some examples of such systems include the human brain \cite{lytton2008computer}, ecosystems \cite{may1977thresholds}, climate systems \cite{robinson2012multistability}, and power grids \cite{machowski2020power}. The desirable states for these systems are generally limited to a single desirable attractor. External disturbances may drive
a system away from the desirable attractor in the phase space, which may cause the system to converge on an undesirable attractor. Thus, studying the response of a system in a desirable state to such perturbations can aid in quantifying the system's stability.

Linear stability analysis, which involves the evaluation of the eigenvalues of the Jacobian matrix at an equilibrium point or the master stability function \cite{pecora1998master}, quantifies the local stability of a system. However, since perturbations affecting systems can be large, linear stability analysis cannot be employed to quantify the stability of a system against finite perturbations. To deal with the effect of finite perturbations, the analytic method of Lyapunov functions \cite{strogatz2018nonlinear} is often used to estimate basins of attractions and thus act as a quantifier of stability. However, Lyapunov functions are extremely hard to construct for many dynamical systems. 

There exist several non-analytical approaches to quantify the stability of a system against finite perturbation \cite{menck2013basin, hellmann2016survivability, mitra2017multiple, kerswell2014optimization, klinshov2015stability, halekotte2020minimal, klinshov2018interval}. A popular measure of stability against finite perturbations known as basin stability \cite{menck2013basin, menck2014dead} relates to the volume of the desirable attractor's basin of attraction. The basin stability of an attractor is defined as the fraction of states that are part of the attractor's basin of attraction in a finite subset of the phase space. Basin stability is numerically estimated using a Monte Carlo simulation. This involves sampling points in a finite region of the phase space and counting the number of points that are part of the attractor's basin of attraction. The ratio of the number of points in the basin of attraction to the total number of points sampled estimates basin stability. A key feature of basin stability is that its computation is numerically tractable for high-dimensional systems, as the standard error associated with the estimation only depends on the number of points sampled and not on the dimension of the system. Basin stability, however, is dependent on a phase space region chosen {\em a priori}. Moreover, basin stability does not say anything about the structure of the basin of attraction or the minimum perturbation that can push the system out of its basin of attraction.

The stability of dynamical systems can also be characterised by the shortest distance from the system's desirable attractor to the boundary of the desirable attractor's basin of attraction \cite{klinshov2015stability, kerswell2014optimization, halekotte2020minimal, walker2004resilience, soliman1989integrity}. This distance is indicative of the minimum perturbation that can push the system out of the basin of attraction of the desirable attractor. There have been a few approaches to numerically estimate this minimum perturbation. One such approach is known as stability threshold \cite{klinshov2015stability}. The stability threshold, however, is only computable for basins with smooth boundaries. Many dynamical systems have fractal basins \cite{aguirre2009fractal}, for which the stability threshold cannot be computed. Another approach presented by R. R. Kerswell et al. \cite{kerswell2014optimization} and L. Halekotte et al. \cite{halekotte2020minimal} makes the use of an optimisation scheme to find the minimum perturbation from an attractor to its basin boundary. However, no criteria exists to determine whether the algorithm has found a local minimum perturbation or the true global minimum perturbation. Furthermore, there is no way of quantifying how much the estimated minimum perturbation deviates from the true minimum perturbation. Due to these reasons, numerically estimating the minimum perturbation that pushes a system out of its basin of attraction may not be reliable.

Instead of knowing the exact value of this minimum perturbation, understanding the perturbation level beyond which the stability of a system gets fairly compromised may provide a good indicator of the system's stability against finite perturbations. To quantify this, we propose a stability quantifier called basin stability bound. This quantifier shares similarities to the manner in which the linear size of a basin of attraction is defined by Delabays et al. \cite{delabays2017size}. Similar to basin stability, a probabilistic sampling-based procedure is used to estimate basin stability bound and approximate the estimation error. This procedure allows for the easy estimation of basin stability bound regardless of the structure of the basin of attraction and the application of basin stability bound to high-dimensional systems. Unlike basin stability, basin stability bound does not directly depend on an {\em a priori} choice of phase space region. Since the phase space of a dynamical system can be unbounded, the basin stability bound is computed in a finite subset of the phase space. However, if the chosen phase space region is large enough, the basin stability bound is independent of this choice.

This paper is organised as follows. We define basin stability bound in section \ref{sec:sec2} and outline the method of its computation. In section \ref{sec:sec3}, we demonstrate the applicability of this stability measure to various dynamical systems. Section \ref{sec:sec4} discusses the applicability of basin stability bound to high-dimensional basins, and section \ref{sec:sec5} concludes the work.

\section{Basin stability bound}\label{sec:sec2}

\subsection{Definition}

Consider an $N$ dimensional dynamical system having a phase space $X$ and at least one finite attractor. The basin stability of a finite attractor $\mathcal{A}$ of the dynamical system is defined as \cite{menck2013basin, menck2014dead}
\begin{equation} \label{eq:eq1}
    S_B (X_P) = \int \chi (x) \: \rho(x, X_P)\: dx
\end{equation}
where $x \in X$, and $X_P$ is a bounded subset of the phase space representing the states to which the system can be perturbed. $\chi (x)$ indicates whether a state is in the basin of attraction $\mathcal{B}$ of the attractor $\mathcal{A}$. $\chi (x)$ is defined as
\begin{equation}
    \chi (x)= 
    \begin{cases}
    1 & \text{if } x \:\in\: \mathcal{B} \\
    0 & \text{otherwise}
    \end{cases}
\end{equation}
$\rho(x, X_P)$ is the probability density of perturbations. It takes non-zero values for $x \in {X_P}$, such that $\int \rho (x, X_P) \: dx = 1$. 

The probability density function $\rho (x, X_P)$ used to define basin stability is typically taken to be a uniform distribution in the region $X_P$ \cite{menck2013basin, menck2014dead}. This implies that every perturbation in the region $X_P$ is equally likely. If the state of the system is represented in Cartesian coordinates, then this density function can be written as
\begin{equation}
    \rho (x, X_P) = 
    \begin{cases}
    {1}/{|X_P|} & \text{if } x \: \in \: X_P\\
    0 & \text{otherwise}
    \end{cases}
\end{equation}
With such a distribution of perturbations, basin stability is defined as
\begin{equation}
     S_B (X_P) = \frac{\text{Vol} (X_P \cap \mathcal{B})}{\text{Vol}(X_P)}
\end{equation}
where $\text{Vol}(X_P \cap \mathcal{B})$ is the volume of the phase space region $X_P \cap \mathcal{B}$, and $\text{Vol}(X_P)$ is the volume of the phase space region $X_P$.

Consider a bounded region ${X_0}\subseteq X$ of the phase space, which represents the extent of perturbations we would like to consider. The basin stability bound is defined in this region. ${X_D}(d)$ is the set of points within a distance $d$ from the attractor that lie in the set $X_0$. It is defined as
\begin{equation}\label{eq:eq3}
    {X_D}(d) = \{x \in {X_0} \:|\: \text{dist}(x,\mathcal{A})<d\}
\end{equation}
where $\text{dist}(x,\mathcal{A})$ is the distance of the state $x$ to the attractor $\mathcal{A}$. If $\bar{d}$ is a distance metric on $X$, then
\begin{equation}
    \text{dist}(x,\mathcal{A}) = \text{inf}\{\bar{d}(x,y)|y \in \mathcal{A} \}
\end{equation}

For all the examples in this paper, we use the Euclidean distance as the distance metric, such that
\begin{equation}
    \bar{d}(x,y) = \sqrt{\sum_{i=1}^{N} a_i ^2(x_i-y_i)^2}
\end{equation}
is the distance between the states $x = (x_1, x_2, ...., x_N)$ and $y = (y_1, y_2, ...., y_N)$. For all examples, we chose $a_i$ to be one with the dimension of $1/x_i$.

Define ${D}$, the set of distances at which the corresponding basin stability is less than a basin stability tolerance $t$. $D$ is written as
\begin{equation}\label{eq:eq8}
    {D} = \{d \in (0, d_\text{max}] \:|\: S_{B} ( {X_D}(d)) < t\}
\end{equation}
where $t \in (0,1]$ is a predefined basin stability tolerance indicating the allowed fraction of undesirable states, and $d_{\text{max}}$ is the maximum distance at which basin stability is computed. We define $d_{\text{max}}$ as 
\begin{equation}\label{eq:eq6}
 d_{\text{max}} \!=\! \text{sup} \{  d \in (0, \infty]| \{x\! \in \! {X} \:|\: \text{dist}(x,\mathcal{A})\!<\!d\} \! \subseteq \! X_0\}.
\end{equation}

The basin stability bound is the minimum distance at which the corresponding basin stability is less than the tolerance $t$. We define the basin stability bound of the attractor $\mathcal{A}$ as
\begin{equation}\label{eq:eq5}
    B_{S} = \begin{cases}
    {\text{inf}({D})}   & \text{if} \: {D} \neq \emptyset  \\
    {d_\text{max}} & \text{otherwise}
    \end{cases}
\end{equation}

Although the basin stability bound is defined in a finite subset $X_0$ of the phase space, its value would not necessarily depend on this choice. The states in the region $X_D (d)$ for $d \in (0,d_\text{max})$ are independent of the choice of $X_0$ and would be the same had $X_D (d)$ been defined using the entire phase space $X$ instead of $X_0$. Thus, if $B_S < d_\text{max}$, then the basin stability bound is independent of the choice of $X_0$.

\subsection{Numerical computation} \label{sec:sec2.2}
A Monte Carlo experiment is used to estimate the basin stability $S_B (X_P)$ \cite{menck2013basin}. To do this, a number of initial conditions, $n$, are sampled from the distribution $\rho (x, {X_P})$ in the region ${X_P}$. If the number of initial conditions that converge to the attractor $\mathcal{A}$ is $n_{A}$, then the estimated basin stability of the attractor is 
\begin{equation}
    \hat{S}_B (X_P) = \dfrac{n_{A}}{n}
\end{equation}

{The basin stability $S_B ({X_P})$, calculated using $n$ trials out of which $n_A$ are considered successes, is the expected probability of success of a Bernoulli experiment. To quantify the uncertainty in the estimate of basin stability, we use the Clopper-Pearson interval (which has coverage guaranteed to be greater than any desired confidence level) \cite{brown2001interval} to provide a $95\%$ confidence interval for the estimated basin stability. With this, the estimated basin stability can be reported within the confidence interval $[\hat{S}_B^L, \hat{S}_B^U] $, where}
\begin{equation}
    {\hat{S}_B^L( n_A,n, \alpha) = \textrm{inf} \{ p \:|\: \mathbb{P}[\textrm{Bin}(n,p) \geq n_A]>\alpha/2 \}}
\end{equation}
{and} 
\begin{equation}
    {\hat{S}_B^U( n_A,n, \alpha) = \textrm{sup} \{ p \:|\: \mathbb{P}[\textrm{Bin}(n,p) \leq n_A]>\alpha/2 \}}
\end{equation}
{Here, $\textrm{Bin}(n,p)$ is a binomial random variable with $n$ number of trials and probability of success $p$, and $\alpha = 0.05$ for a $95 \%$ confidence interval.}

{The expression of $\hat{S}_B^L$ and $\hat{S}_B^U$, with $\alpha = 0.05$, can be also written as}
\begin{equation}
    {\hat{S}_B^L(n_A,n) = B(0.025, n_A, n - n_A +1 )}
\end{equation}
{and}
\begin{equation}
    {\hat{S}_B^U(n_A,n) = B(0.975, n_A+1, n - n_A )}
\end{equation}
{where $B$ is the quantile function of the beta distribution. The confidence interval associated with the estimated basin stability is independent of the dimension of the system, making basin stability easily computable for high-dimensional systems. }

Assuming the basin stability $\hat{S}_B(X_D (d))$ is computed $ \forall d\in(0,d_\text{max}]$ using $n$ samples for every basin stability computation, then, the basin stability bound can be computed using equation \eqref{eq:eq8} and equation \eqref{eq:eq5}. {At the distance $d$, the corresponding estimate of basin stability with a confidence level of 0.95 is reported in the confidence interval $[\hat{S}_B^L(X_D(d)), \hat{S}_B^U(X_D(d))]$.} Because of the statistical uncertainty in estimating basin stability, the estimate of basin stability bound is also subject to a degree of uncertainty. The basin stability bound computed using the lower bound of the confidence interval of the estimated basin stability provides a conservative estimate of the basin stability bound, whereas the basin stability bound computed using the upper bound of the confidence interval of the estimated basin stability provides a less conservative estimate of the basin stability bound. These two basin stability bound values can serve as the lower and upper bound of a confidence interval representing the uncertainty in the computation of basin stability bound that stems from the uncertainty in the basin stability computation. To quantify the uncertainty in computing the basin stability bound, we have assumed that the basin stability is computed $\forall d \in (0, d_\text{max}]$. However, it is only numerically possible to compute the basin stability at discrete values of $d$. This marginally adds to the basin stability bound's computational error. Appendix \ref{sec:appA} describes the detailed computation procedure, computational efficiency and computational error of basin stability bound.

{At a distance arbitrarily close to the attractor, the basin stability is always one. The lower bound of the confidence interval of an estimated basin stability of $1$ is $\hat{S}_B^L( n,n) = B(0.025, n, 1 )
$. If $B(0.025, n, 1 )<t$, then the minimum distance at which the lower bound of the confidence interval of basin stability is less than $t$ does not exist. Thus, if $B(0.025, n, 1 )<t$, then the lower bound of the confidence interval of the estimated basin stability bound does not exist. Hence, it is necessary that $B(0.025, n, 1 ) \geq t$ for the numerical estimation of the basin stability bound to be tractable. Since $B(0.025, n, 1 )$ increases with an increase in $n$, the higher the value of $t$, the larger the number of sampled points required for the basin stability bound computation. When $t$ tends to $1$, the number of sampled points $n$ needed for $B(0.025, n, 1 ) \geq t$ to be satisfied tends to infinity. Hence, in this case, numerical errors are not tractable, and thus, this sampling procedure cannot be used to compute the basin stability bound for $t=1$.}

A higher value of $t$ would indicate that a lesser number of undesirable states are allowed within the basin stability bound. For most purposes, one would ideally like a value of $t$ close to one, implying that very few undesirable states are within the basin stability bound. However, for computing the basin stability bound, the larger the value of $t$, the more the number of samples required for its computation. To balance computational efficiency and reliability, we have chosen $t= 0.95$ and $n = 300$ for the examples in this paper, thus satisfying the {criterion $B(0.025, n, 1 ) \geq t$}. This value of $t=0.95$ indicates that at most $5\%$ of undesirable states are within the basin stability bound.

\section{Results}\label{sec:sec3}

\subsection{Amazonian vegetation model}
As the first example, we examine an ecological model of the Amazon rainforest \cite{menck2013basin, mitra2015integrative, da2001savanna, hirota2011global}. This model shows two stable states, one being a fertile forest state and the other being a barren savanna state. The following differential equation describes the model
\begin{equation}
    \frac{dC}{dt} = \begin{cases}
    r(1-C)C -xC & \text{if } \: C > C_{\text{crit}}\\
    -yC & \text{if} \: C < C_{\text{crit}}
    \end{cases}
\end{equation}
$C$ is the relative forest cover. $C$ grows at a rate $r$ and dies at a rate $x$, when $C$ is greater than the critical forest cover $C_\text{crit}$. $C$ dies at a rate $y$, when  $C< C_{\text{crit}}$. The higher the aridity $A$ of the region, the more the critical forest cover. We set $C_{\text{crit}} = A$.

This system has two fixed points, $C_F = 1-{x}/{r}$, corresponding to the stable forest state, and $C_S = 0$, corresponding to the stable savanna state. The stable forest state exists if $C_F>C_\text{crit}$, and the stable savanna state exists if $C_\text{crit}>0$. Factors such as global warming can affect the aridity of the rainforest, thus changing the critical forest cover. If the system is in the forest state, an increase in aridity can push the system to a bifurcation point, after which the forest state ceases to exist. As the aridity increases, the border dividing the basin of attraction of the forest state and the basin of attraction of the savanna state moves towards the stable forest state until it collides with the forest state, causing the forest state to vanish.

The local stability of the forest state fails to detect the shrinking of the basin of attraction captured in the state's basin stability \cite{menck2013basin}. However, basin stability fails to capture the gradual loss in stability of the system near the bifurcation point. When the forest state is close to the bifurcation point, small perturbations in the direction of the basin boundary can push the system into the savanna state. Basin stability bound captures this aspect.

The parameters chosen for the differential equation are $r = 1$, $x=0.5$ and $y = 1$. {The} basin stability and basin stability bound {of the forest state} are computed in the phase space region $[0,1]$. We use $n=300$ points sampled from a uniform distribution for every basin stability computation. {When the forest state does not exist, the forest's state basin stability and basin stability bound are considered zero.} 

\begin{figure}[h]
    \centering
    \includegraphics[scale=0.465]{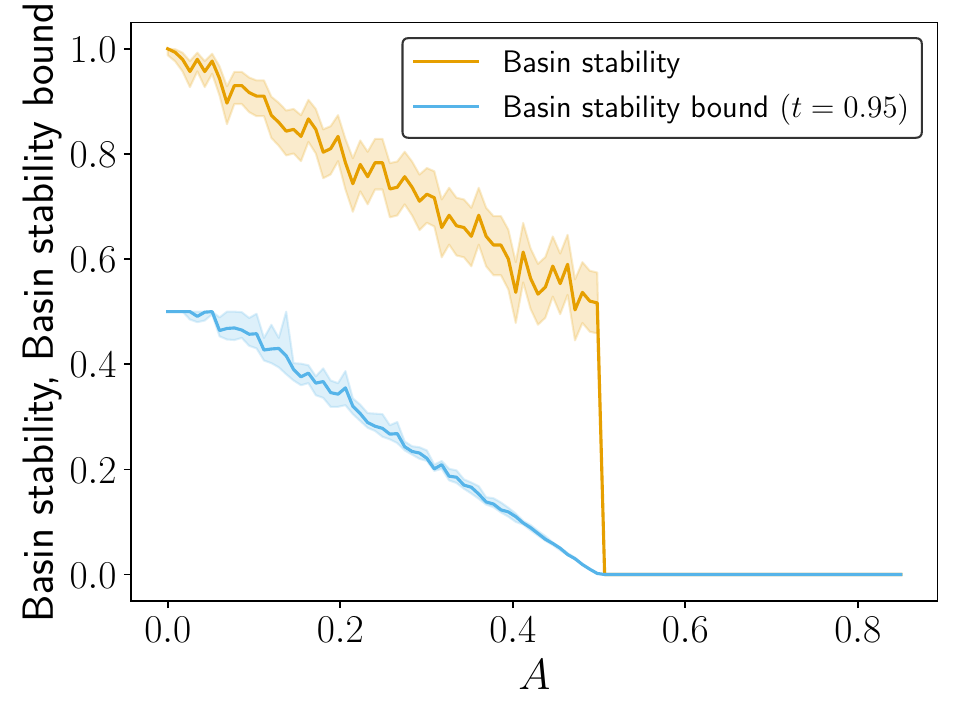}
    \caption{Stability of the forest state. Basin stability and basin stability bound plotted versus aridity, $A$. The shaded region represents the {95\% confidence interval}.}
    \label{fig:figure1}
\end{figure}

Fig.~\ref{fig:figure1} shows that basin stability bound captures the gradual loss of stability near the bifurcation point of the forest state, whereas basin stability fails to capture this.

\begin{figure}[h]
    \centering
    \includegraphics[scale=0.465]{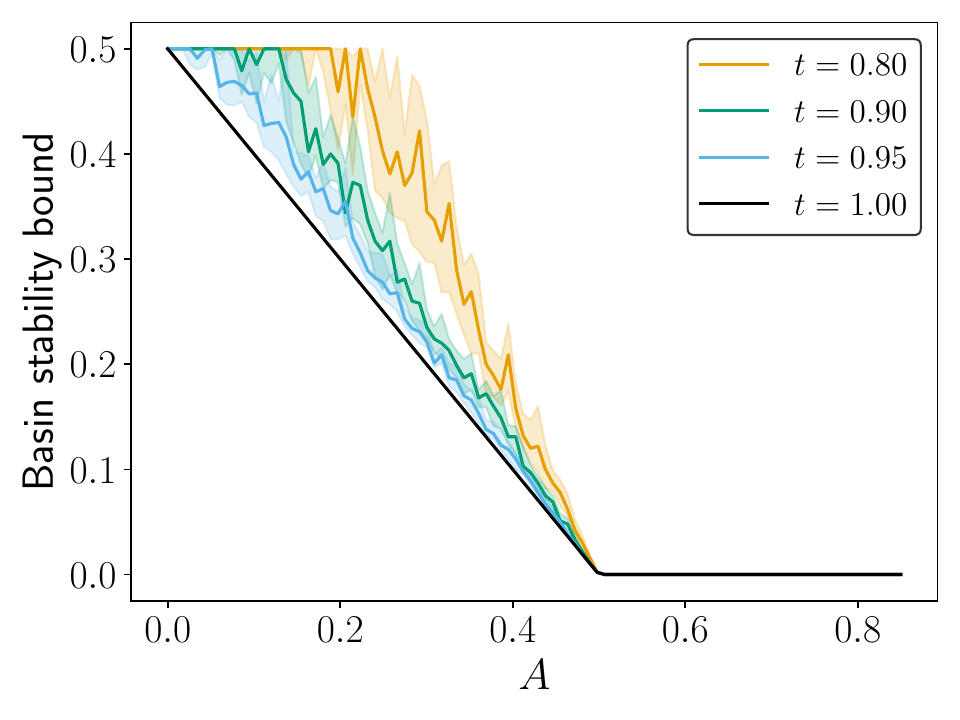}
    \caption{Basin stability bound of the forest state versus Aridity  $A$, for different values of the tolerance $t$. The shaded region represents the {95\% confidence interval}.}
    \label{fig:figure2}
\end{figure}

In Fig.~\ref{fig:figure2}, the basin stability bound of the forest is plotted for different values of the tolerance $t$. The basin stability bound for $t=1$ is equal to the minimum distance from the attractor to the basin boundary, which is given as $C_F - C_\text{crit}$. It is observed that for all four tolerance values, the gradual loss of stability of the forest state is captured by the basin stability bound.

\subsection{Ship capsize model}
Waves and wind forces constantly buffet ships at sea, and under some conditions, ships may capsize. It is interesting to find the ship's stability against such perturbations.

The motion of a ship can be defined in six degrees of freedom \cite{rawson2001basic}. We are concerned here with the motion of a ship in one specific degree of freedom: the rotation of a vessel about its longitudinal axis, which runs horizontally through the length of the ship. The ship's motion in this degree of freedom is called roll, which is responsible for capsizes.  

The roll motion of a ship at sea can be modelled as \cite{thompson1990ship}
\begin{equation}
    \Ddot{x}+\beta \dot{x}+c(x)=F \sin (\omega t)
\end{equation}
where $x$ is the scaled roll angle of the ship, $t$ is the scaled time, $\beta $ is the damping coefficient of the ship, $c$ is restoring moment of bouyancy forces. Forcing due to wind and waves in regular beam seas is assumed to be sinusoidal, with amplitude $F$ and angular frequency $\omega$.

In high winds, capsizing towards the wind is discounted. The restoring force in such conditions is modelled as $c(x) = x - x^2$. The restoring force $c(x)$ can be expressed in terms of a potential $v(x) = \frac{1}{2} x^2 - \frac{1}{3}x^3$, such that $c(x)=dv(x)/dx$. This potential has a local maxima at $x=1$. Escape over the local maxima of this potential corresponds to capsize.

The equations describing the rolling motion of a ship in high winds can be written as
\begin{subequations}
\begin{eqnarray}
     \dot{x} &=& y\\
    \dot{y} &=& - \beta \dot{x}-x+x^2+F \sin (\omega t)
\end{eqnarray}
\end{subequations}

These differential equations have been extensively studied before \cite{thompson1987fractal, thompson1989chaotic, thompson1989basin}. The dynamical system described by the differential equations can have multiple attractors, all of which are desirable. The set of initial conditions that converge to the attractors of the system is known as the system's safe basin. Thus, the safe basin describes the set of all states that are safe for the system and do not lead to capsizing.

With an increase in the forcing amplitude $F$, the safe basin gets eroded by fractal incursions, thus compromising the system's stability. This erosion of the safe basin of a system is known as basin erosion. To understand how basin erosion affects the stability of a system, various integrity measures that quantify the stability of such systems have been proposed \cite{soliman1989integrity, thompson2019dynamical}. These integrity measures include the global integrity measure, which is the normalised volume of the safe basin, and the local integrity measure, which is the minimum distance from the attractor to the boundary of the safe basin in the Poincar\'e section. The global and local integrity measures are computed using a grid of points in the phase space. This numerical approach to compute the integrity measures is not robust, as the accuracy of the computed integrity measures is not quantifiable. 

To reliably measure the stability of this system, we define basin stability and basin stability bound for the safe basin of the system. The basin stability of the safe basin is the fraction of states in a finite phase space region $X_P$ that are part of the safe basin $\mathcal{B'}$. It is given by
\begin{equation}\label{eq:eq16}
     S_B (X_P) = \frac{\text{Vol} (X_P \cap \mathcal{B'})}{\text{Vol}(X_P)}
\end{equation}
The basin stability bound is defined from the attractor in the Poincar\'e section using the definition of basin stability in equation (\ref{eq:eq16}).

The parameters chosen for the differential equation are $\beta = 0.$1 and $\omega = 0.85$. An initial phase $\phi = \pi$ was chosen for the forcing term, such that the initial forcing is $-F$. The basin stability of the safe basin is computed in the region $X_P = [-0.8,1.2] \times [-1,1]$. The attractor corresponding to what would be observed physically under the slow increase of $F$ from zero is considered \cite{thompson1990ship, soliman1989integrity}. The Poincar\'e section at phase $\phi = \pi$ is considered. Basin stability bound is computed in the region $X_0 = \{z \in X| \text{dist}(z,z_0) < 1\}$, where $X$ is the phase space, $z_0$ is the attractor in the Poincar\'e section at phase $\phi = \pi$, and $\text{dist}(z,z_0)$ is the Euclidean distance between $z = (x,y)$ and $z_0 = (x_0,y_0)$ defined as $\text{dist}(z,z_0) = \sqrt{a^2(x-x_0)^2+b^2(y-y_0)^2}$. $a$ and $b$ are both taken to be one with the dimensions of $1/x$ and $1/y$, respectively. We use $n=300$ points sampled from a uniform distribution for every basin stability computation.

\begin{figure}[h]
    \centering
    \includegraphics[scale=0.465]{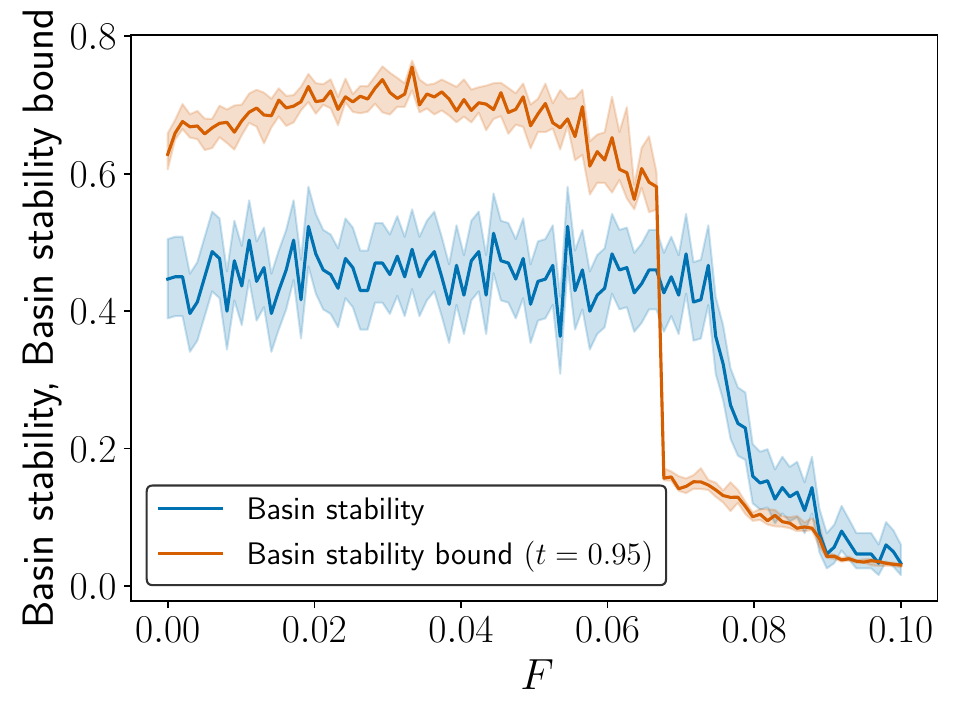}
    \caption{Stability of a ship against capsize. Basin stability and basin stability bound plotted versus the forcing amplitude, $F$. The shaded regions around the plots represent the {95\% confidence interval}.}
    \label{fig:figure3}
\end{figure}

In Figure \ref{fig:figure3}, both the basin stability and the basin stability bound show a decrease in their values due to erosion of the safe basin on increasing the forcing amplitude $F$. Thompson et al.  \cite{thompson1990ship} observed that stability is compromised around $F \approx 0.07$. Around this value, the basin stability bound shows a sharp fall due to the attractor shifting close to the basin boundary. Basin stability does not capture this, thus highlighting the utility of basin stability bound as a measure of the system's integrity.

\subsection{Power grids}
Power grids facilitate the generation and consumption of electricity over vast geographical regions. During normal functioning, a power grid operates in its stable synchronous state, in which all parts of the grid function at the same frequency \cite{machowski2020power}. Power grids can be subject to many internal and external perturbations that can push the system out of its synchronous state. These perturbations can lead to a cascading failure, in which a small initial failure triggers a string of further failures \cite{buldyrev2010catastrophic, motter2002cascade, schafer2018dynamically}. Such failures can propagate through the power grid and destabilise large parts of the grid. Cascading failures are devastating as they can result in large-scale power blackouts that can leave millions of people without electricity \cite{ind_blackout, turkey_blackout, italy_blackout}. Thus, it is vital to identify vulnerabilities in power grids in order to prevent future grid failures.

We use a complex network power grid representation to model power grids, with nodes as generators and consumers and edges as transmission lines. Generators and consumers are modelled using the swing equation \cite{machowski2020power}. The equations that describe the dynamics of the grid are \cite{filatrella2008analysis, menck2014dead}
\begin{subequations} \label{eq:eq14}
    \begin{eqnarray}
    \dot{\theta_i} &=& \omega_i\\
    \dot{\omega_i} &=& -\alpha_i \omega_i + P_i - \sum^N_{j=1} K_{ij}\sin(\theta_i - \theta_j)
\end{eqnarray}
\end{subequations}
where, for the $i\:^\text{th}$ node,\\
$\theta_i$ is the phase angle in the frame rotating at the synchronous grid frequency, \\ $\omega_i$ is the angular velocity in the frame rotating at the synchronous grid frequency,\\
$\alpha_i$ is the damping factor,\\ $P_i$ is the net power generated or consumed.\\ $K_{ij}$ is the transmission capacity between node $i$ and node $j$, provided they are connected to each other. If node $i$ and node $j$ are not connected, then $K_{ij} = 0$. 

The fixed point of equation (\ref{eq:eq14}) corresponds to the stable synchronous state of the grid. In this state, the $i\:^\text{th}$ node has the phase $\theta_i^s$ and frequency $0$. Several undesirable non-synchronous states also exist to which perturbations can push the grid to \cite{filatrella2008analysis,menck2014dead, chiang2011direct}. To study the effect of finite perturbations at individual nodes in a grid, a variant of basin stability known as single-node basin stability has been introduced \cite{menck2014dead}. The single-node basin stability of node $i$ in the network is the basin stability conditioned on perturbations only hitting the node $i$ (perturbations to $\theta_i$ and $\omega_i$) from the initial synchronous state. If single-node perturbations occur in the region $X_{sn}$, such that $(\theta_i, \omega_i) \in X_{sn}$ is a single-node perturbation at the $i^\text{th}$ node, then, the single-node basin stability of node $i$ is the basin stability with perturbations conditioned in the region
\begin{equation}\label{eq:eq18}
\begin{split}
X_i^0 = &\{ (\theta, \omega) \in X | (\theta_i, \omega_i) \in X_{sn} \\ & \land (\forall j \neq i: \theta_j = \theta^s_j \land \omega_j = 0) \}
\end{split}
\end{equation}
where $\theta = (\theta_1, \theta_2,..., \theta_N)$ and $\omega = (\omega_1, \omega_2,..., \omega_N)$.

The 2-D slices of the basin of attraction corresponding to single-node perturbations are convoluted and fractal \cite{halekotte2021transient}. Thus, single-node basin stability may not be a reliable stability quantifier as it does not consider the basin's structure. Complementary to single-node basin stability, we define a variant of basin stability bound called single-node basin stability bound. The single-node basin stability bound of node $i$ is the basin stability bound with perturbations conditioned to the node $i$ from the initial synchronous state. If single-node perturbations occur in the region $X_{sn}$, then $X_i^0$ represents the perturbation space for computing the single-node basin stability bound of node $i$. 

To understand how network structure affects single-node stability, we compute the single-node basin stability and the single-node basin stability bound for every node of several synthetic power grid networks. For this, 200 Erd\"os–R\'enyi networks, each with 40 nodes and 54 edges, were generated. In each network, half of the nodes were taken to be generators, and half of the nodes were taken to be consumers. The following parameters were used: $K=8$ for every transmission line, $\alpha = 0.1$ for every node, $P=1$ for every generator, and $P=-1$ for every consumer. We choose $X_{sn} = [-\pi, \pi] \times [-100,100]$. The single-node basin stability of node $i$ is computed in the region $X_i^0$ given by equation (\ref{eq:eq18}). The single-node basin stability bound of node $i$ is computed in the region $\{x \in X_{i}^0| \text{dist}(x,\mathcal{A})< 15\}$, where $\mathcal{A}$ is the synchronous state of the grid. Most basin stability bound values are less than 15, and increasing the region for basin stability bound computation does not change these values. Thus, a larger region for computing basin stability bound is not chosen. The distance between a state in the phase space to the attractor is defined as $\text{dist}(x,\mathcal{A}) = \sqrt{\sum_{i=0}^N [a^2(\theta_i-\theta_i^s)^2+b^2\omega_i^2]}$. $a$ and $b$ are both taken to be one with the dimensions of $1/\theta_i$ and $1/\omega_i$, respectively. The basin stability bound is computed with a tolerance $t = 0.95$, and $n=300$ points sampled from a uniform distribution are used for every basin stability computation.

There have been several studies on the impact of network topology on the single-node basin stability of power grids \cite{menck2014dead, kim2016building, nitzbon2017deciphering,schultz2014detours}. It has been observed that based on single-node basin stability, nodes inside dead ends and dead trees are likely to be less stable \cite{menck2014dead}. The stability of such network motifs can be better understood by studying the relation between single-node stability and a network centrality measure known as betweenness centrality. Betweenness centrality measures the importance of a node in a network based on how many shortest paths pass through it. The betweenness centrality of node $i$ in a network is defined as \cite{freeman1977set}
\begin{equation}    b_i = \sum_{j\neq i, k\neq i,k>j} \dfrac{\sigma_{jk}^{i}}{\sigma_{jk}}
\end{equation}
where $\sigma_{jk}^{i}$ is the number of shortest paths from node $j$ to node $k$ which pass through node $i$, and $\sigma_{jk}$ is the number of shortest paths which pass from node $j$ to node $k$.

\begin{figure}[h]
     \centering
     \begin{subfigure}{0.5\textwidth}
         \centering
         \caption{}
         \includegraphics[scale = 0.465]{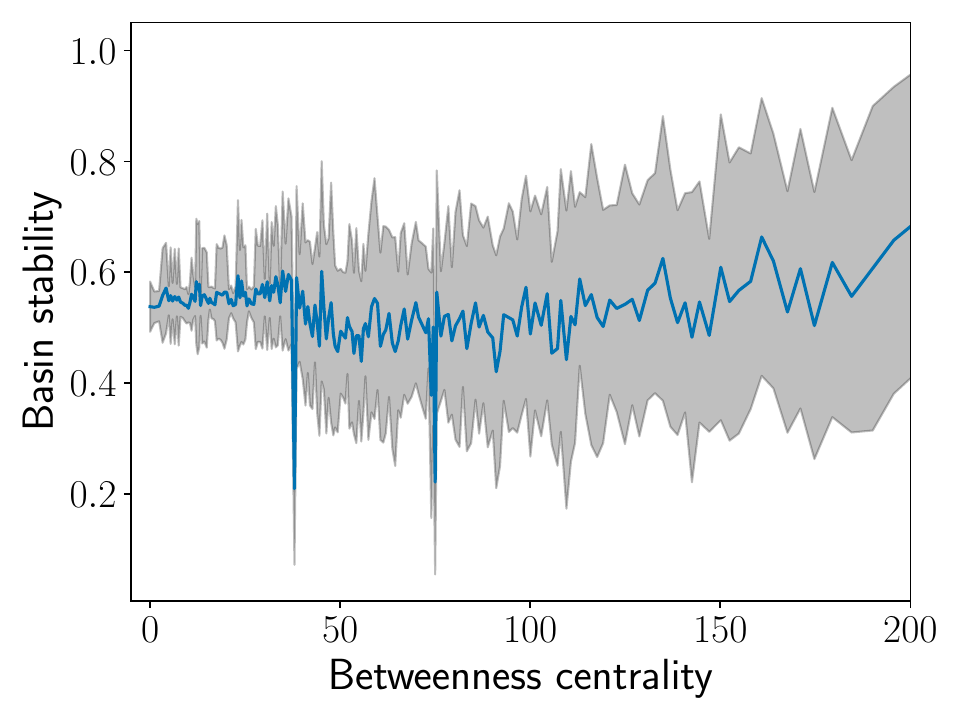}
         \label{fig:fig4a}
     \end{subfigure}
     \hfill
     \begin{subfigure}{0.5\textwidth}
         \centering
         \caption{}
         \includegraphics[scale = 0.465]{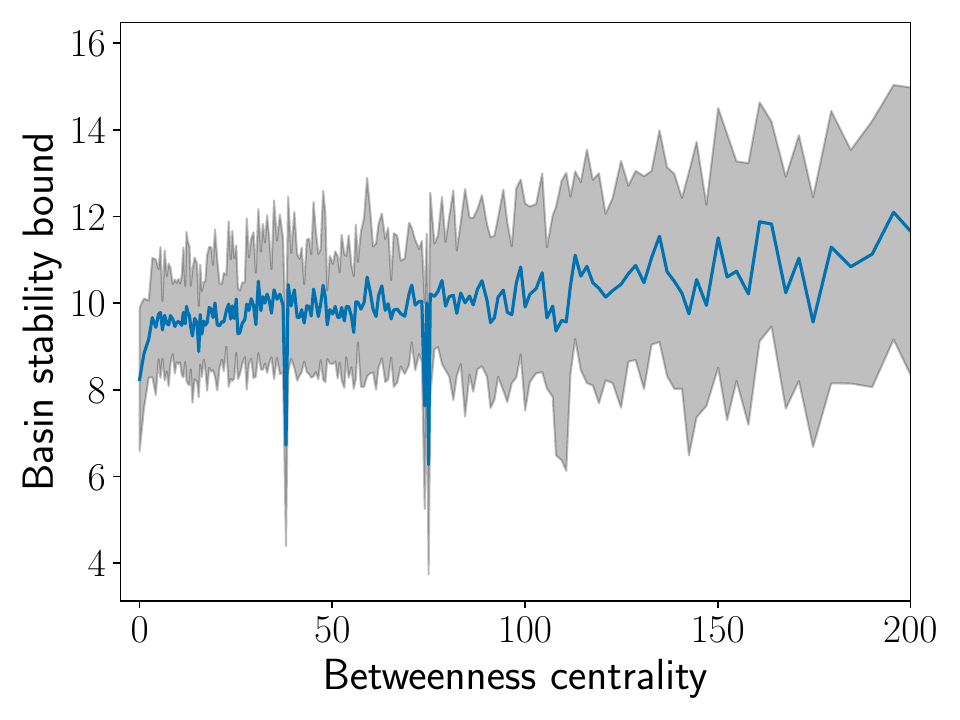}
         \label{fig:fig4b}
     \end{subfigure}
        \caption{The dependence of single-node stability on betweenness centrality. (a) Single-node basin stability versus betweenness centrality. (b) Single-node basin stability bound versus betweenness centrality. The blue line shows the average trend, and the grey region indicates the {basin stability bound values within} $\pm 1$ standard deviation.}
        \label{fig:fig4}
\end{figure}

In Fig.~\ref{fig:fig4a}, marked dips in the single-node basin stability are observed corresponding to betweenness centrality values of $N-2$, $2N-6$, and $2N-5$, where $N$ is the number of nodes in the networks (here, $N=40$). These dips correspond to nodes inside dead ends and dead trees \cite{menck2014dead}. In Fig.~\ref{fig:fig4b}, dips in the single-node basin stability bound are observed at the same betweenness centrality values as the dips of single-node basin stability in Fig.~\ref{fig:fig4a}. In addition to these dips, a marked dip in the single-node basin stability bound is observed at a betweenness centrality of $0$. This dip corresponds to dead-end nodes and contains over $20 \%$ of the nodes in all the generated networks, whereas every other dip contains less than $0.1 \%$ of nodes. Like single-node basin stability, single-node basin stability bound characterises nodes inside dead ends and dead trees as less stable. However, unlike single-node basin stability, single-node basin stability bound characterises dead ends as less stable. Since a large percentage of nodes in power grids are dead ends, the fact that these nodes have low basin stability bound values could prove useful for identifying power grid vulnerabilities and preventing future blackouts.

\section{Application to high-dimensional basins}\label{sec:sec4}
In the three examples in section \ref{sec:sec3}, we examined low-dimensional basins using a uniform distribution of perturbations to compute the basin stability and the basin stability bound. However, for high dimensional basins, such as that of a plant-pollinator network \cite{bastolla2009architecture, halekotte2020minimal} or a power grid \cite{filatrella2008analysis,menck2014dead}, a uniform distribution of perturbations may not provide useful information about the basin of attraction as most of the sampled states would be located towards the surface of the region used for sampling, with only a few states located near the attractor. This would imply that stronger perturbations are disproportionately more sampled than moderate and weaker perturbations, thus, making such a distribution unideal.

To account for this, we consider a distribution of states such that the distances of the states from the attractor are uniformly distributed within a chosen phase space region. With this distribution, sampling a state at any distance from the attractor is equally likely. Furthermore, using this distribution, the number of sampled states required for estimating basin stability bound does not depend on the dimension of the system (appendix \ref{sec:appA}). This allows basin stability bound to be easily computable for high-dimension systems. 

As an example, we consider a network of second-order Kuramoto oscillators described by the following differential equations
\begin{subequations} \label{eq:eq20}
    \begin{eqnarray}
    \dot{\theta_i} &=& \omega_i\\
    \dot{\omega_i} &=& 0.1 \omega_i  \pm 1 - K\sum^N_{j=1} A_{ij}\sin(\theta_i - \theta_j)
\end{eqnarray}
\end{subequations}
where $A_{ij}$ is the network's adjacency matrix, and an equal number of oscillators have values $-1$ and $+1$. 

We consider an Erd\"os-Reyni network of 40 nodes and 54 edges. 
The $40$ dimension basin slice $X$ formed by $\{\omega_1, \omega_2, ..., \omega_n\}$ is considered. 

The density function for perturbations whose distances from the attractor are uniformly distributed can be written as
\begin{equation}
    \rho (r, X_P) = 
    \begin{cases}
    {1}/{|X_P|} & \text{if } r \: \in \: X_P\\
    0 & \text{otherwise}
    \end{cases}
\end{equation}
where $r$ is the state of the system in spherical coordinates.

We consider a large phase space region given by $X_0 = \{x \in X | \text{dist}(x, \mathcal{A})<500\}$, where $\text{dist}(x, \mathcal{A})$ is the Eucledean distance between $x = (\omega_1, \omega_2,..., \omega_n)$ and the attractor $\mathcal{A} = (0, 0, ..., 0)$ in the basin slice $X$. The basin stability and the basin stability bound are computed in the region $X_0$. We use $n=300$ sampled states from the density function $\rho(r,X_P)$ for every basin stability computation. Tolerance $t = 0.95$ is used to compute the basin stability bound. The stability of the system is studied by varying the coupling $K$.

\begin{figure}[h]
     \centering
     \begin{subfigure}{0.5\textwidth}
         \centering
         \caption{}
         \includegraphics[scale = 0.465]{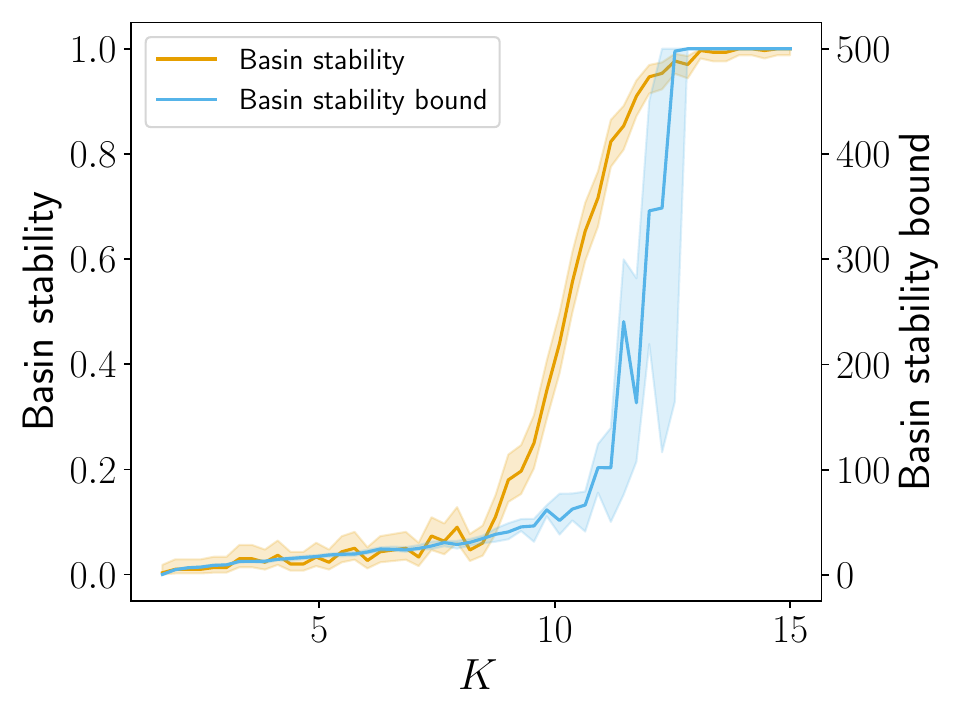}
         \label{fig:fig5a}
     \end{subfigure}
     \hfill
     \begin{subfigure}{0.5\textwidth}
         \centering
         \caption{}
         \includegraphics[scale = 0.465]{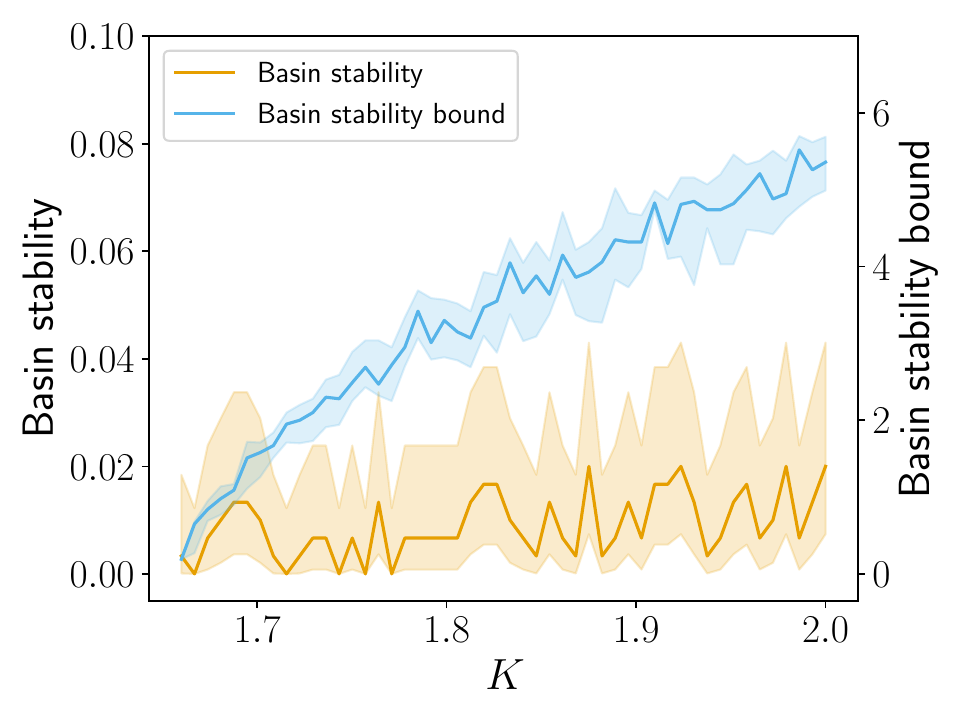}
         \label{fig:fig5b}
     \end{subfigure}
        \caption{The dependence of basin stability and basin stability bound on the coupling $K$. The shaded region indicates the {95\% confidence interval}.}
        \label{fig:fig5}
\end{figure}

In Fig. \ref{fig:fig5a}, both basin stability and basin stability bound increase on increasing the coupling $K$. 

Fig \ref{fig:fig5b} shows a zoomed-in version of the plot in Fig. \ref{fig:fig5a} for low values of the coupling $K$. Basin stability fails to indicate an increasing trend and has a very large relative estimation error associated with it {as the number of sampled points $n_A$ that converge to the attractor are far less than the total number of sampled points $n$}. Conversely, the basin stability bound distinctively increases on increasing $K$ and the basin stability bound estimation error scales with the {estimated} basin stability bound value. This makes basin stability bound far more useful than basin stability when the size of the basin of attraction is unknown, and the {\em a priori} phase space region for computing basin stability cannot be chosen suitably. 

\section{Conclusion}\label{sec:sec5}

We have introduced a novel distance-based probabilistic stability quantifier for dynamical systems and have shown its applicability to real-world systems. 

In the case of the Amazon rainforest model, we have observed how the basin stability bound provides early warning signs before the loss of stability of the forest state. These early warning signs are not observed in the basin stability of the forest state. In the ship capsize model, the basin stability bound provides a much better indicator of an impending catastrophe.

In power grids, we have successfully introduced basin stability bound to quantify the single-node stability of power grids. We have found that the single-node basin stability bound identifies vulnerabilities in a large class of nodes that have not been previously detected as vulnerable according to single-node basin stability. To complement basin stability, we believe that basin stability bound can employed to examine power grid stability.

Lastly, we have also demonstrated the applicability of basin stability bound to high-dimensional basins and have shown its utility over basin stability.

\section{Acknowledgements}
We are thankful to Arnab Acharya for the valuable discussions. S. Banerjee acknowledges the J C Bose
National Fellowship provided by SERB, Government of India,
Grant No. JBR/2020/000049.

\begin{appendices}

\section{}\label{sec:appA}

To compute the basin stability bound, a finite region of the phase space, $X_0$, is first considered. The basin stability $\hat{S}_B({X_D} (d))$ is computed using $n$ sampled points from $d = d_\text{max}$ to $d = d_0$ with a finite step size $\delta$. $X_D (d)$ is defined in equation \eqref{eq:eq3}, $d_\text{max}$ is defined in equation \eqref{eq:eq6}, and $d_0$ is the largest distance such that $\hat{S}_B({X_D} (d_0))=1$. For the basin stability estimation $\hat{S}_B({X_D} (d-\delta))$, sampled points from the previous basin stability estimation $\hat{S}_B({X_D} (d))$ that fall in the region $X_D(d-\delta)$, are reused. This significantly reduces the total number of sampled points required for the computation of basin stability bound and allows one to choose a very small value of $\delta$.

In this study, we consider two different density functions of perturbations for computing basis stability --- a density function $\rho_1 (x, X_P)$ representing perturbations uniformly distributed in the phase space $X_P$, and a density function $\rho_2 (x, X_P)$ representing perturbations in the phase space $X_P$ with their distances from the attractor distributed uniformly, such that sampling a state at any distance is equally likely.

For the density function $\rho_1 (x, X_P)$, the basin stability is the fraction of the volume of the basin of attraction in the region $X_P$. If basin stability is computed in the region $X_D(d)$, then the fraction of the number of sampled points that fall in the region $X_D(d-\delta)$ is proportional to $\text{Vol}(X_D(d-\delta)) / \text{Vol}(X_D(d))$, which is further proportional to $((d-\delta)/d)^N$, where $N$ is the dimension of the system. Thus, the fraction of points that can be reused for the next basin stability computation would be very low for high-dimensional systems.

On the other hand, for the density function $\rho_2$, if basin stability has been computed in the region $X_D(d)$, then the fraction of the number of sampled points that fall in the region $X_D(d-\delta)$ is proportional to $(d-\delta)/d$. Thus, the number of sampled points that can be reused for every successive basin stability computation does not depend on the dimension of the system, and the total number of sampled points required for basin stability bound computation is independent of the dimension of the system.

Between two successive distances $d+\delta$ and $d$, the basin stability $S_B( {X_D} (d'))$ for $d' \in (d,d+\delta)$ is not known. If the basin stability $S_B ( {X_D} (d))$ is known, then for $d' \in [d, d+\delta]$, $S_B (X_D(d')) \geq S_B(X_D(d)) \left( {d}/{(d+\delta)} \right)^N$ for density function $\rho_1$ and $S_B (X_D(d')) \geq S_B(X_D(d)) \left( {d}/{(d+\delta)} \right)$ for density function $\rho_2$. {Thus,} the worst-case estimate of basin stability for $d' \in [d,d+\delta]$ is 
\begin{equation}
    {\hat{S}'_B(X_D(d), X_D(d+\delta)) =
    \hat{S}_B(X_D(d)) \left( {d}/{(d+\delta)} \right)^K }
\end{equation}
{where $K=N$ for density function $\rho_1$ and $K=1$ for density function $\rho_2$. The basin stability bound is numerically computed using this worst-case basin stability estimate.}

If the estimated basin stability $\hat{S}_B({X_D} (d_\text{max})) < t$, the value $d_\text{max}$ is added a set $D$.  For $d \in \{d_0, d_0+\delta,....,d_\text{max} -\delta \}$, if the worst-case basin stability estimate $\hat{S}'_B (X_D(d), X_D(d+\delta))<t$, the values of $d$ are noted and added to a set ${D}$. The basin stability bound can then be calculated using equation \eqref{eq:eq5}. 

Stemming from the uncertainty in estimating basin stability (section \ref{sec:sec2.2}), the worst-case basin stability estimate {$\hat{S}'_B(X_D(d), X_D(d+\delta))$} can be reported in the confidence interval {$$[\hat{S}'^L_B (X_D(d), X_D(d+\delta)),\hat{S}'^U_B (X_D(d), X_D(d+\delta))],$$ where
\begin{equation}
    \hat{S}'^L_B (X_D(d), X_D(d+\delta)) = \hat{S}^L_B(X_D(d)) \left( {d}/{(d+\delta)} \right) ^K
\end{equation}
and
\begin{equation}
    \hat{S}'^U_B (X_D(d), X_D(d+\delta)) = \hat{S}^U_B(X_D(d)) \left( {d}/{(d+\delta)} \right) ^K
\end{equation}
with $K = N$ for the density function $\rho_1$ and $K= 1$ for the density function $\rho_2$.} 

Following how the basin stability bound confidence interval is defined in section \ref{sec:sec2.2}, we report a more conservative estimate of this confidence interval using the worst-case basin stability estimate. {Thus,} the basin stability bound computed using the lower (upper) bound of the confidence interval of the worst-case basin stability estimate corresponds to the lower (upper) bound of the confidence interval of the estimated basin stability bound. The difference in the worst-case basin stability estimate $\hat{S}'_B (X_D(d), X_D(d+\delta))$ and the basin stability estimate $\hat{S}_B(X_D(d))$ is proportional to {$1-(d/(d+\delta))^K$ ($K=N$ for $\rho_1$ and $K=1$ for $\rho_2$)}. Due to the resampling of points, $\delta$ can be chosen to be very small, and, in our examples, it is always the case that {$\hat{S}^U_B-\hat{S}^L_B >>1-(d/(d+\delta))^K$}. This makes the error contribution from the finite step size $\delta$ {negligible compared to} the error from the statistical uncertainty in computing basin stability. 




\end{appendices}


\bibliographystyle{unsrt}
\bibliography{bibliography.bib}

\begin{thebibliography}{10}

\bibitem{feudel2018multistability}
Ulrike Feudel, Alexander~N Pisarchik, and Kenneth Showalter.
\newblock Multistability and tipping: From mathematics and physics to climate and brain—minireview and preface to the focus issue.
\newblock {\em Chaos: An Interdisciplinary Journal of Nonlinear Science}, 28(3), 2018.

\bibitem{lytton2008computer}
William~W Lytton.
\newblock Computer modelling of epilepsy.
\newblock {\em Nature Reviews Neuroscience}, 9(8):626--637, 2008.

\bibitem{may1977thresholds}
Robert~M May.
\newblock Thresholds and breakpoints in ecosystems with a multiplicity of stable states.
\newblock {\em Nature}, 269(5628):471--477, 1977.

\bibitem{robinson2012multistability}
Alexander Robinson, Reinhard Calov, and Andrey Ganopolski.
\newblock Multistability and critical thresholds of the {G}reenland ice sheet.
\newblock {\em Nature Climate Change}, 2(6):429--432, 2012.

\bibitem{machowski2020power}
Jan Machowski, Zbigniew Lubosny, Janusz~W Bialek, and James~R Bumby.
\newblock {\em Power system dynamics: stability and control}.
\newblock John Wiley \& Sons, 2020.

\bibitem{pecora1998master}
Louis~M Pecora and Thomas~L Carroll.
\newblock Master stability functions for synchronized coupled systems.
\newblock {\em Physical review letters}, 80(10):2109, 1998.

\bibitem{strogatz2018nonlinear}
Steven~H Strogatz.
\newblock {\em Nonlinear dynamics and chaos with student solutions manual: With applications to physics, biology, chemistry, and engineering}.
\newblock CRC press, 2018.

\bibitem{menck2013basin}
Peter~J Menck, Jobst Heitzig, Norbert Marwan, and J{\"u}rgen Kurths.
\newblock How basin stability complements the linear-stability paradigm.
\newblock {\em Nature physics}, 9(2):89--92, 2013.

\bibitem{hellmann2016survivability}
Frank Hellmann, Paul Schultz, Carsten Grabow, Jobst Heitzig, and J{\"u}rgen Kurths.
\newblock Survivability of deterministic dynamical systems.
\newblock {\em Scientific reports}, 6(1):29654, 2016.

\bibitem{mitra2017multiple}
Chiranjit Mitra, Anshul Choudhary, Sudeshna Sinha, J{\"u}rgen Kurths, and Reik~V Donner.
\newblock Multiple-node basin stability in complex dynamical networks.
\newblock {\em Physical Review E}, 95(3):032317, 2017.

\bibitem{kerswell2014optimization}
RR~Kerswell, Chris~CT Pringle, and AP~Willis.
\newblock An optimization approach for analysing nonlinear stability with transition to turbulence in fluids as an exemplar.
\newblock {\em Reports on Progress in Physics}, 77(8):085901, 2014.

\bibitem{klinshov2015stability}
Vladimir~V Klinshov, Vladimir~I Nekorkin, and J{\"u}rgen Kurths.
\newblock Stability threshold approach for complex dynamical systems.
\newblock {\em New Journal of Physics}, 18(1):013004, 2015.

\bibitem{halekotte2020minimal}
Lukas Halekotte and Ulrike Feudel.
\newblock Minimal fatal shocks in multistable complex networks.
\newblock {\em Scientific reports}, 10(1):11783, 2020.

\bibitem{klinshov2018interval}
Vladimir~V Klinshov, Sergey Kirillov, J{\"u}rgen Kurths, and Vladimir~I Nekorkin.
\newblock Interval stability for complex systems.
\newblock {\em New Journal of Physics}, 20(4):043040, 2018.

\bibitem{menck2014dead}
Peter~J Menck, Jobst Heitzig, J{\"u}rgen Kurths, and Hans Joachim~Schellnhuber.
\newblock How dead ends undermine power grid stability.
\newblock {\em Nature communications}, 5(1):3969, 2014.

\bibitem{walker2004resilience}
Brian Walker, Crawford~S Holling, Stephen~R Carpenter, and Ann Kinzig.
\newblock Resilience, adaptability and transformability in social--ecological systems.
\newblock {\em Ecology and society}, 9(2), 2004.

\bibitem{soliman1989integrity}
M.~S. Soliman and J.~M.~T. Thompson.
\newblock Integrity measures quantifying the erosion of smooth and fractal basins of attraction.
\newblock {\em Journal of Sound and Vibration}, 135(3):453--475, 1989.

\bibitem{aguirre2009fractal}
Jacobo Aguirre, Ricardo~L Viana, and Miguel~AF Sanju{\'a}n.
\newblock Fractal structures in nonlinear dynamics.
\newblock {\em Reviews of Modern Physics}, 81(1):333, 2009.

\bibitem{delabays2017size}
Robin Delabays, Melvyn Tyloo, and Philippe Jacquod.
\newblock The size of the sync basin revisited.
\newblock {\em Chaos: An Interdisciplinary Journal of Nonlinear Science}, 27(10), 2017.

\bibitem{brown2001interval}
Lawrence~D Brown, T~Tony Cai, and Anirban DasGupta.
\newblock Interval estimation for a binomial proportion.
\newblock {\em Statistical science}, 16(2):101--133, 2001.

\bibitem{mitra2015integrative}
Chiranjit Mitra, J{\"u}rgen Kurths, and Reik~V Donner.
\newblock An integrative quantifier of multistability in complex systems based on ecological resilience.
\newblock {\em Scientific reports}, 5(1):16196, 2015.

\bibitem{da2001savanna}
Leonel Da~Silveira Lobo~Sternberg.
\newblock Savanna--forest hysteresis in the tropics.
\newblock {\em Global Ecology and Biogeography}, 10(4):369--378, 2001.

\bibitem{hirota2011global}
Marina Hirota, Milena Holmgren, Egbert~H Van~Nes, and Marten Scheffer.
\newblock Global resilience of tropical forest and savanna to critical transitions.
\newblock {\em Science}, 334(6053):232--235, 2011.

\bibitem{rawson2001basic}
Kenneth~John Rawson and Eric~Charles Tupper.
\newblock {\em Basic Ship Theory Volume 1}, volume~1.
\newblock Butterworth-Heinemann, 2001.

\bibitem{thompson1990ship}
John Michael~Tutill Thompson, R.~C.~T. Rainey, and M.~S. Soliman.
\newblock Ship stability criteria based on chaotic transients from incursive fractals.
\newblock {\em Philosophical Transactions of the Royal Society of London. Series A: Physical and Engineering Sciences}, 332(1624):149--167, 1990.

\bibitem{thompson1987fractal}
J.~M.~T. Thompson, S.~R. Bishop, and L.~M. Leung.
\newblock Fractal basins and chaotic bifurcations prior to escape from a potential well.
\newblock {\em Physics Letters A}, 121(3):116--120, 1987.

\bibitem{thompson1989chaotic}
John Michael~Tutill Thompson.
\newblock Chaotic phenomena triggering the escape from a potential well.
\newblock {\em Proceedings of the Royal Society of London. A. Mathematical and Physical Sciences}, 421(1861):195--225, 1989.

\bibitem{thompson1989basin}
J.~M.~T. Thompson and Y~Ueda.
\newblock Basin boundary metamorphoses in the canonical escape equation.
\newblock {\em Dynamics and stability of systems}, 4(3-4):285--294, 1989.

\bibitem{thompson2019dynamical}
J~Michael~T Thompson.
\newblock Dynamical integrity: three decades of progress from macro to nanomechanics.
\newblock In Stefano Lenci, Giuseppe Rega, et~al., editors, {\em Global nonlinear dynamics for engineering design and system safety}, volume 588, pages 1--26. Springer, 2019.

\bibitem{buldyrev2010catastrophic}
Sergey~V Buldyrev, Roni Parshani, Gerald Paul, H~Eugene Stanley, and Shlomo Havlin.
\newblock Catastrophic cascade of failures in interdependent networks.
\newblock {\em Nature}, 464(7291):1025--1028, 2010.

\bibitem{motter2002cascade}
Adilson~E Motter and Ying-Cheng Lai.
\newblock Cascade-based attacks on complex networks.
\newblock {\em Physical Review E}, 66(6):065102, 2002.

\bibitem{schafer2018dynamically}
Benjamin Sch{\"a}fer, Dirk Witthaut, Marc Timme, and Vito Latora.
\newblock Dynamically induced cascading failures in power grids.
\newblock {\em Nature communications}, 9(1):1975, 2018.

\bibitem{ind_blackout}
Central Electricity Regulatory~Commission (CERC).
\newblock Report on the grid disturbances on 30th {J}uly and 31st {J}uly 2012.
\newblock Technical report, 2012.

\bibitem{turkey_blackout}
{Project Group Turkey}.
\newblock Report on blackout in {T}urkey on 31st {M}arch 2015.
\newblock Technical report, European Network of Transmission System Operators for Electricity (ENTSO-E), 2015.

\bibitem{italy_blackout}
{Union for the Coordination of Transmission of Electricity (UCTE)}.
\newblock Final report of the investigation committee on the 28th {S}eptember 2003 blackout in {I}taly.
\newblock Technical report, European Network of Transmission System Operators for Electricity (ENTSO-E), 2003.

\bibitem{filatrella2008analysis}
Giovanni Filatrella, Arne~Hejde Nielsen, and Niels~Falsig Pedersen.
\newblock Analysis of a power grid using a {K}uramoto-like model.
\newblock {\em The European Physical Journal B}, 61:485--491, 2008.

\bibitem{chiang2011direct}
Hsiao-Dong Chiang.
\newblock {\em Direct methods for stability analysis of electric power systems: theoretical foundation, BCU methodologies, and applications}.
\newblock John Wiley \& Sons, 2011.

\bibitem{halekotte2021transient}
Lukas Halekotte, Anna Vanselow, and Ulrike Feudel.
\newblock Transient chaos enforces uncertainty in the {B}ritish power grid.
\newblock {\em Journal of Physics: Complexity}, 2(3):035015, 2021.

\bibitem{kim2016building}
Heetae Kim, Sang~Hoon Lee, and Petter Holme.
\newblock Building blocks of the basin stability of power grids.
\newblock {\em Physical Review E}, 93(6):062318, 2016.

\bibitem{nitzbon2017deciphering}
Jan Nitzbon, Paul Schultz, Jobst Heitzig, J{\"u}rgen Kurths, and Frank Hellmann.
\newblock Deciphering the imprint of topology on nonlinear dynamical network stability.
\newblock {\em New Journal of Physics}, 19(3):033029, 2017.

\bibitem{schultz2014detours}
Paul Schultz, Jobst Heitzig, and J{\"u}rgen Kurths.
\newblock Detours around basin stability in power networks.
\newblock {\em New Journal of Physics}, 16(12):125001, 2014.

\bibitem{freeman1977set}
Linton~C Freeman.
\newblock A set of measures of centrality based on betweenness.
\newblock {\em Sociometry}, pages 35--41, 1977.

\bibitem{bastolla2009architecture}
Ugo Bastolla, Miguel~A Fortuna, Alberto Pascual-Garc{\'\i}a, Antonio Ferrera, Bartolo Luque, and Jordi Bascompte.
\newblock The architecture of mutualistic networks minimizes competition and increases biodiversity.
\newblock {\em Nature}, 458(7241):1018--1020, 2009.

\end{thebibliography}

\end{document}